\documentclass{article}

\usepackage{spconf_modified,amsmath,epsfig}
\usepackage{amssymb}
\usepackage{amsthm}
\usepackage{color}
\usepackage{url}
\usepackage{cite}
\usepackage{mdframed}

\usepackage{algorithmicx}
\usepackage{algpseudocode}

\newcommand{\vect}[1]{\mathbf{#1}}

\newcommand{\maximize}[1]{{\underset{{#1}}{\mathrm{maximize}}}}

\theoremstyle{remark}

\newtheorem{definition}{Definition}

\newtheorem{property}{Property}

\title{Multi-Objective Signal Processing Optimization: \\ The Way to Balance Conflicting Metrics in 5G Systems}

\name{Emil Bj{\"o}rnson, Eduard Jorswieck, M{\'e}rouane Debbah, and Bj{\"o}rn Ottersten}

\address{}
\begin{document}
\maketitle

\begin{abstract}
The evolution of cellular networks is driven by the dream of ubiquitous wireless connectivity:  \emph{Any data service is instantly accessible everywhere}. With each generation of cellular networks, we have moved closer to this wireless dream; first by delivering wireless access to voice communications, then by providing wireless data services, and recently by delivering a WiFi-like experience with wide-area coverage and user mobility management. The support for high data rates has been the main objective in recent years \cite{Tombaz2011a}, as seen from the academic focus on sum-rate optimization and the efforts from standardization bodies to meet the peak rate requirements specified in IMT-Advanced. In contrast, a variety of metrics/objectives are put forward in the technological preparations for 5G networks: higher peak rates, improved coverage with uniform user experience, higher reliability and lower latency, better energy efficiency, lower-cost user devices and services, better scalability with number of devices, etc. These multiple objectives are coupled, often in a conflicting manner such that improvements in one objective lead to degradation in the other objectives. Hence, the design of future networks calls for new optimization tools that properly handle the existence and tradeoffs between multiple objectives.

In this article, we provide a review of multi-objective optimization (MOO), which is a mathematical framework to solve design problems with multiple conflicting objectives \cite{Zadeh1963a,Marler2004a,Branke2008a,Bot2009a,Bjornson2013d}. In contrast to conventional heuristic approaches where some objectives are converted into constraints, MOO enables a rigorous network design. MOO has been applied in many engineering and economic related fields, but has received little attention from the signal processing and wireless communication communities. We provide a survey of the basic definitions, properties, and algorithmic tools in MOO.  This reveals how signal processing algorithms are used to visualize the inherent conflicts between 5G performance objectives, thereby allowing the network designer to understand the possible operating points and how to balance the objectives in an efficient and satisfactory way. For clarity, we provide a case study on massive multiple-input multiple-output (MIMO) systems, which is one of the key enablers of 5G cellular networks.

\end{abstract}

\vspace{-3mm}

\section*{\small INTRODUCTION}

\vspace{-2mm}

We are currently at a point in time when many researchers in industry and academia are trying to formalize their expectations and requirements on the next generation wireless communication networks. These views are expressed in various magazine articles, white papers, and plenary talks. To get a sense of the range of expectations, one can take a look at the project \emph{Mobile and wireless communications Enablers for the Twenty-twenty Information Society} (METIS), \url{http://www.metis2020.com/}, where telecommunications manufacturers, network operators, and academic partners are gathering their 5G requirements. The following summarizes their main objectives \cite{Osseiran2013a}:
\begin{itemize}

\item Higher user data rates: $10$--$100$ times higher average user rates are expected, at least in urban scenarios.

\item Higher area data rates: $1000$ times higher average rates per unit area are anticipated.

\item More connected devices: With the respective expected increases in user and area rates, $10$--$100$ times more devices can be accommodated per unit area.

\item Higher energy efficiency (EE): The throughput should be improved without increasing the operational cost or the energy consumption, thus greatly improving the EE. If EE is measured as area data rate per power expenditure, this requires a $1000$ times EE improvement.

\end{itemize}

Furthermore, \emph{heterogeneity} appears as a keyword that can be tied to a variety of network aspects:

\begin{itemize}

\item Heterogeneous networks: The combination of access points with different ranges, traffic loads, radio access technologies, licensed/unlicensed spectrum, and hardware capabilities makes the network highly heterogeneous. The same deployment strategy cannot be used everywhere and the same resource management scheme cannot be used throughout the day.

\item Heterogeneous user conditions: As the performance requirements become tighter, the mobility and pathloss of a specific user determines its quality-of-service, unless the network is designed to counteract these effects.

\item Heterogeneous devices: The differences in functionality and hardware capability of user devices are expected to grow. Large handheld devices can, for example, achieve high data rates by spatial multiplexing and advanced signal processing, while small sensors seek low data rates under extremely tight energy constraints.

\item Heterogeneous service requirements: Some cyber-physical systems and public-safety applications require very fast and reliable response times, while best-effort delivery is fine for other types of data services. Similarly, certain multimedia applications have tight and continuous quality-of-service requirements, while other services are bursty in nature.

\end{itemize}

There are apparently many different requirements, or \emph{objectives}, to keep in mind when designing future wireless networks. Unfortunately, these objectives cannot be treated separately because they are coupled; sometimes in a consistent fashion, but often in conflicting ways such that improvements in one objective lead to deterioration of other objectives. This is because the same network \emph{resources} (e.g., time, frequency, space, power, and hardware)  play key roles in all these requirements/objectives, but in incompatible ways. As a simple example, higher peak user rates can be achieved by using more power (which affects the EE), allocating more transmission resources to users with good channels (which means less uniform user experience and higher latencies), or making use of intricate signal processing algorithms (which increases the complexity and cost of user devices).

In order to achieve the ambitious 5G goals, efficient network operation with respect to \emph{all} the conflicting 5G objectives is required. This calls for a design framework that handles multiple objectives and supports the search for the best attainable operating point. But can we really formulate and solve multi-objective problems rigorously or is heuristic trial-and-error the only option? Is there even any optimal solution? These are questions that we address in this article.

\vspace{-1mm}

\subsection*{\small\em CONVENTIONAL SINGLE-OBJECTIVE OPTIMIZATION}

\vspace{-1mm}

The conventional approach to physical-layer system optimization is that of selecting a scalar \emph{network utility function}  that is maximized under a set of constraints \cite{Kelly1997a,Palomar2006a}. A common problem formulation is that of maximizing the weighted sum of the users' data rates under transmit power constraints \cite{Weingarten2006a,Liu2011a,Bjornson2013d}. Alternatively, one can minimize the transmitted power under the constraint of guaranteeing certain data rates to each user \cite{Rashid1998a,Chiang2008a}. In recent years, the EE (in bit/Joule) has also arisen as a utility function \cite{Chen2011a,Isheden2012a,Bjornson2015a}.

In essence, the conventional approach is to select one of the objectives listed above as the sole objective, while the other objectives are transformed into constraints. The inherent heuristic assumptions are: 1) one of the objectives is of dominating importance; and 2) it is known beforehand what are good values for the constraints related to the other objectives. Moreover, the short-term values of the different objectives are usually considered in these network utility problems and not the long-term values which are of main importance in the network design. Given the increased complexity due to heterogeneity, the need for long-term network optimization, and the diverse expectations on 5G networks, the conventional approach is no longer viable. However, we show later how to construct more appropriate single-objective problems.

\vspace{-2mm}

\section*{\small NEW PARADIGM: MULTI-OBJECTIVE OPTIMIZATION}

\vspace{-2mm}

Instead of assuming that one of the objectives is the sole objective, the fundamental approach is to recognize the existence of multiple objectives \cite{Zadeh1963a}: $g_1(\vect{x}), g_2(\vect{x}), \ldots, g_M(\vect{x})$ where $M$ is the number of objectives. These objective functions can, for example, be area throughput, guaranteed rates for different classes of users, number of simultaneous users, energy efficiency, etc. Explicit examples are given later in this article, while the theory is applicable for any arbitrary functions. The notation $\vect{g}(\vect{x}) = [g_1(\vect{x}), g_2(\vect{x}), \ldots, g_M(\vect{x})]^T$ is used to emphasize that the objective is vector-valued.

The available resources (e.g., time, frequency, space, power, and hardware) are modeled by a compact set $\mathcal{X} \subset \mathbb{R}^{D}$, which is called the \emph{resource bundle} and has any finite dimension $D$. Each vector $\vect{x} \in \mathcal{X}$ represents a feasible way of utilizing the network resources. The satisfaction of this resource utilization equals $g_m(\vect{x}) \in \mathbb{R}$ with respect to the $m$th objective function. A larger value corresponds to higher satisfaction. For tractability we assume that $g_m(\vect{x})$ is a bounded continuous function of $\vect{x}$ and non-negative. We also assume that it exists a point $\vect{x}_0 \in \mathcal{X}$ such that $g_m(\vect{x}_0) = 0$ for all $m$. This operating point is the dissatisfaction of turning off the network and makes the satisfaction (for each objective) become a number from zero and upwards. Not all practical objectives satisfy these conditions by nature; for example, latency and error probability are typically to be minimized. However, there are standard transformations that reformulate such metrics into objective functions in our framework  \cite{Marler2004a,Branke2008a,Bot2009a,Bjornson2013d}.

A key assumption is that the $M$ objectives are \emph{not} ordered and therefore studied without any preconceptions---all doors are kept open. In contrast to game theory, where each objective belongs to one of the competing agents, we assume that there is a \emph{network designer} that would like to design the network to maximize \emph{all} the $M$ objectives simultaneously:
\begin{align} \label{eq:optimization-problem}
\maximize{\vect{x}} & \quad \vect{g}(\vect{x})=[g_1(\vect{x}), g_2(\vect{x}), \ldots, g_M(\vect{x})]^T \\
\mathrm{subject} \, \mathrm{to} & \quad \vect{x} \in \mathcal{X}. \notag
\end{align}
Note that \eqref{eq:optimization-problem} is the maximization of the vector $\vect{g}(\vect{x})$ containing the $M$ objectives, which is defined as maximizing all elements simultaneously. This is known as a \emph{multi-objective optimization problem (MOOP)} or, alternatively, as a multi-criteria or vector optimization problem \cite{Zadeh1963a,Marler2004a,Branke2008a,Bot2009a,Bjornson2013d}. These types of problems arise in many engineering fields because of the difficulty to find a scalar metric that exactly describes what we would like to achieve. We review the main concepts and properties related to MOOPs in this article. We provide the basic tools to understand the structure of MOOPs and how to solve these problems in practice. The properties are stated without proofs, while we recommend \cite{Marler2004a,Branke2008a,Bot2009a} for further details and \cite{Bjornson2013d} for a recent survey aimed at communication applications.

\begin{property} \label{property:global-optimum}
The $M$ objectives in \eqref{eq:optimization-problem} are conflicting and since there is no total order of vectors, there is (generally) no global optimum to the MOOP in \eqref{eq:optimization-problem}.
\end{property}

\begin{figure}
	\begin{center}
		\includegraphics[width=\columnwidth]{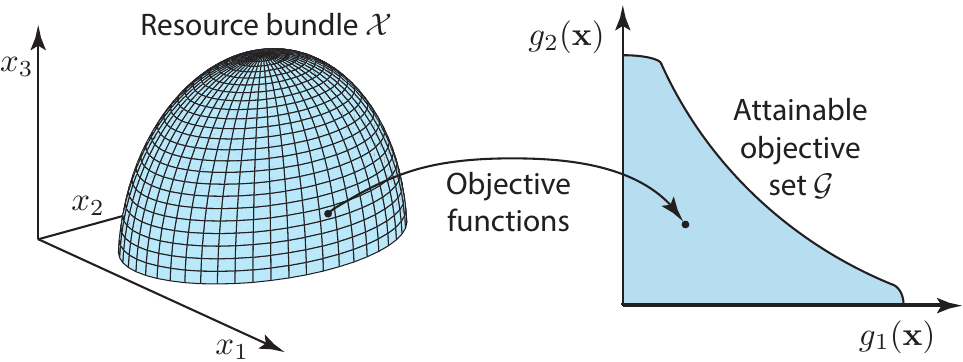} \vskip-3mm
	\caption{Illustration of a MOOP with a three-dimensional resource bundle $\mathcal{X}$ and a two-dimensional attainable objective set. For each resource utilization $\vect{x}= [x_1 \, x_2 \, x_3]^T \in \mathcal{X}$, the objective functions $g_1(\vect{x})$ and $g_2(\vect{x})$ assign a vector $\vect{g}(\vect{x}) \in \mathcal{G}$.}
	\label{fig:set-example}
	\end{center} \vskip-6mm
\end{figure}

This is the first important insight from the multi-objective framework; we cannot solve \eqref{eq:optimization-problem} in any globally optimal sense because there are \emph{only subjectively optimal solutions}. Therefore, we turn the attention to the attainable objective set
\begin{equation}
\mathcal{G} = \{  \vect{g}(\vect{x}) : \vect{x} \in \mathcal{X} \}
\end{equation}
which contains all the combinations of objective values $g_1(\vect{x}), g_2(\vect{x}), \ldots, g_M(\vect{x})$ that are simultaneously attainable under the available resources. The relationship between the resource bundle $\mathcal{X}$ and the attainable objective set $\mathcal{G}$ is visualized in Fig.~\ref{fig:set-example}. Note that the origin is always in the objective set, $\vect{0} = [0 \, \ldots \, 0]^T \in \mathcal{G}$, due to the assumptions above.

When formulating the MOOP, the resource bundle $\mathcal{X}$ is selected to minimize the preconditions made on the utilization of network resources. This keeps all the options open, because it is generally difficult to articulate the network requirements \emph{a priori}---at least in a strict mathematical sense. Nevertheless, the resource bundle can include certain fundamental network performance constraints (e.g., that the $M$ metrics should be better than in previous network generations).

\vspace{-2mm}

\subsection*{\small\em PARETO OPTIMAL OPERATING POINTS}

The shape of the attainable objective set $\mathcal{G}$ depends on the objective functions and the resource bundle $\mathcal{X}$, but it is usually a compact set with the property that $\vect{g} \in \mathcal{G}$ implies $c \vect{g} \in \mathcal{G}$ for all $ c \in [0,1]$ (i.e., the performance can be uniformly degraded). The set $\mathcal{G}$ can be convex or non-convex. Although Property \ref{property:global-optimum} expresses that there is no global optimum, most points in $\mathcal{G}$ are strictly suboptimal. In fact, any point in the interior of $\mathcal{G}$ can be discarded because there exist other points in $\mathcal{G}$ that are more preferable with respect to all $M$ objectives. The remaining points belong to the Pareto boundary.

\begin{definition}[Pareto boundary]
The strong Pareto boundary, $\partial \mathcal{G}$, consists of all points $\vect{g} \in \mathcal{G}$ for which there does not exist any $\vect{g}' \in \mathcal{G} \setminus \{ \vect{g} \}$ with $g'_m \geq g_m$ for $m=1,\ldots,M$.
\end{definition}

The strong Pareto boundary consists of the attainable operating points that cannot be objectively dismissed, because none of the objectives can be improved without degrading other objectives. Evidently, any point that is not on the strong Pareto boundary is suboptimal because there exist other operating points that are better or at least as good for \emph{every} objective. The strong Pareto boundary is as close to global optimality as one can get in multi-objective optimization; the operating points in $\partial \mathcal{G}$ are mutually unordered and can only be compared by subjective means. Each point $\vect{g} \in \partial \mathcal{G}$ describes a particular tradeoff between the $M$ objectives. Hence, the Pareto boundary describes the set of (Pareto) efficient potential operating points from which we, as network designers, should select the one that is subjectively preferable to us.

\begin{figure}
	\begin{center}
		\includegraphics[width=\columnwidth]{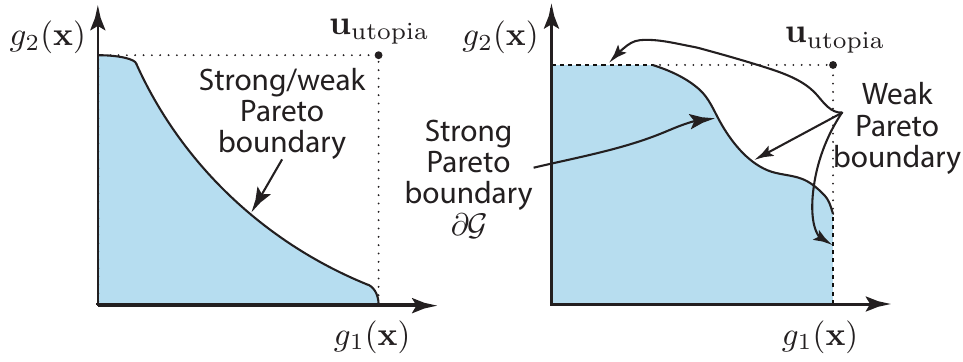} \vskip-3mm
	\caption{Illustration of the Pareto boundary, which is either the complete upper boundary of $\mathcal{G}$ (left) or a subset of the upper boundary (right). The unattainable utopia point is also shown.}
	\label{fig:set-example2}
	\end{center} \vskip-6mm
\end{figure}

The strong Pareto boundary is a subset of the upper boundary of  $\mathcal{G}$. The complete  upper boundary is referred to as the \emph{weak} Pareto boundary and also contains points were some of the objectives (but not all) can be improved without degrading other objectives. This is illustrated in Fig.~\ref{fig:set-example2}, where the strong Pareto boundary either equals the complete upper boundary (left set) or is a strict subset thereof (right set). Fig.~\ref{fig:set-example2} also shows the \emph{utopia point}, which is defined as
\vspace{-6mm}

\begin{equation} \label{eq:utopiapoint}
\vect{u}_{\mathrm{utopia}}  = [u_1 \, \ldots u_M]^T = \begin{bmatrix}
\max_{\vect{x} \in \mathcal{X}} f_1(\vect{x}) \\
\vdots \\
\max_{\vect{x} \in \mathcal{X}} f_M(\vect{x})
      \end{bmatrix}.
\end{equation}
\vspace{-3mm}

\noindent This is the ideal operating point that simultaneously maximizes all $M$ objectives. If $\vect{u}_{\mathrm{utopia}} \in \mathcal{G}$, the MOOP is trivial because the strong Pareto boundary consists of only the utopia point, $\partial \mathcal{G} = \{ \vect{u}_{\mathrm{utopia}} \}$, and it is the unique global optimum.

\begin{property} \label{property:global-optimum2}
Any MOOP with multiple conflicting objective functions is nontrivial in the sense that $\vect{u}_{\mathrm{utopia}} \not \in \mathcal{G}$ and, consequently, there is no global optimum.
\end{property}

Single-objective optimization problems are MOOPs with $M=1$ and are thus trivial from the MOO perspective.

Since the Pareto boundary consists of all tentative effective operating points, we need to find the network parameters (i.e., the resource utilizations) that attain these points.

\begin{definition}[Pareto optimal point]
A point $\vect{x}^* \in \mathcal{X}$ in the resource bundle is a Pareto optimal point if $\vect{g}(\vect{x}^*) \in \partial \mathcal{G}$.
\end{definition}

The mapping from a Pareto optimal point $\vect{x}^*$ to the Pareto boundary is given by the vector-valued multi-objective function $\vect{g}(\vect{x}^*)$ and is, hopefully, given in closed form. The inverse mapping is, on the other hand, hard to derive in most cases. The multi-objective function might not be bijective, which means that multiple points in $\mathcal{X}$ can give exactly the same objective point. This happens frequently when transmitting from multi-antenna arrays, where the beamforming coefficients are only unique up to a common phase rotation \cite{Bjornson2013d}.

\subsection*{\small\em SOLVING A MOOP BY VISUALIZATION}

In practice, we would like to go beyond the Pareto boundary and actually \emph{solve} the MOOP, in the sense of selecting a \emph{single} Pareto optimal point $\vect{x}^*$ and its corresponding operating point $\vect{g}(\vect{x}^*) \in \partial \mathcal{G}$. To this end, we need to bring in the subjective preference of the network designer to compare different operating points at the Pareto boundary. This is not as simple as it might seem, because neither the Pareto boundary $\partial \mathcal{G}$ nor the objective set $\mathcal{G}$ are known beforehand. Simple closed-form expressions are seldom available. In fact, one needs to spend considerable computational resources on learning the objective set. For example, one can characterize $\mathcal{G}$ by computing a discrete set of sample points, which enables the network designer to visualize the different possibilities and make an informed decision. This is known as the \emph{a posteriori method}, because the network designer formulates its subjective preference after the numerical computations have taken place \cite{Zadeh1963a}.

We describe two approaches to compute sample points:

\begin{enumerate}

\item Traverse the resource bundle $\mathcal{X}$ by computing $\vect{g}(\vect{x})$ over a finite grid of $\vect{x} \in \mathcal{X}$. For example, if $0 \leq x_m \leq 1$ then we can limit ourselves to the 6 discrete values $x_m \in \{ 0, \, 0.2, \, 0.4, \, 0.6, \, 0.8, \, 1\}$. If the same number of discrete values are taken for all $D$ resource variables in $\mathcal{X}$, we have $6^D$ grid points to consider.

\item Traverse the strong Pareto boundary $\partial \mathcal{G}$ by searching for the outermost point in $\mathcal{G}$ in different directions. The search directions can, for example, be represented by vectors $\vect{v} = [v_1 \, \ldots v_M]^T$ that point out (nonnegative) geometric directions from the origin (recall that $\vect{0} \in \mathcal{G}$ by definition). Each search corresponds to solving the single-objective optimization problem
\begin{align} \label{eq:line-search-problem}
\maximize{\vect{x},\lambda} & \quad \lambda \\ \notag
\mathrm{subject} \, \mathrm{to} & \quad g_m(\vect{x}) \geq  \lambda v_m, \quad m=1,\ldots,M, \\ \notag
 & \quad \vect{x} \in \mathcal{X},
\end{align}
which is a referred to as a weighted Chebyshev problem in the MOO literature \cite{Marler2004a,Branke2008a,Bot2009a,Bjornson2013d} (in fact, it is the epigraph form of it \cite{Boyd2004a}). If $\lambda^*$ is the optimal value for a given $\vect{v}$, we can be sure that $\lambda^* \vect{v} \in \mathcal{G}$ and that this point lies on the weak Pareto boundary (upper boundary). If needed, one can guarantee to attain the strong Pareto boundary by slightly modifying \eqref{eq:line-search-problem}; see \cite{Marler2004a}. By solving \eqref{eq:line-search-problem} for a finite set of search directions (e.g., equally spaced in the angular sense), one can obtain a set of sample points that characterizes the weak/strong Pareto boundary.
\end{enumerate}

These two approaches have their respective pros and cons. The first approach is computationally efficient, assuming that the function values $\vect{g}(\vect{x})$ are easy to evaluate. The main limiting factor might be the memory storage, since the number of samples scales exponentially with $D$. Extensive postprocessing might also be required because most sample points will be in the interior of $\mathcal{G}$ and can be discarded since there are other samples that are better with respect to all $M$ objectives. The resource bundle can sometimes be parameterized more efficiently by exploiting the objective functions. This can be used to improve the resolution of the objective set $\mathcal{G}$ using fewer samples. For example, transmit beamforming can be represented by one parameter per user \cite[Section 3.2]{Bjornson2013d}, which removes redundancy in multi-antenna wireless communications where the number of beamforming coefficients equals the number of users \emph{times} the number of transmit antennas.

The second approach guarantees a high resolution because every sample point lies on the weak Pareto boundary. The downside is the computational complexity, which is proportional to the complexity of solving the search problem in \eqref{eq:line-search-problem}. Indeed, this approach can only be utilized if there is a tractable way of solving \eqref{eq:line-search-problem}. This is the case whenever there exists an efficient way to make a membership test; that is, to determine if a given point $\tilde{\vect{g}} \in \mathbb{R}^M$ belongs to the objective set or not. We elaborate on this in the box ``Finding the Pareto Boundary by Bisection'' below.

\begin{figure} \vskip-2mm
	\begin{center}
		\includegraphics[width=\columnwidth]{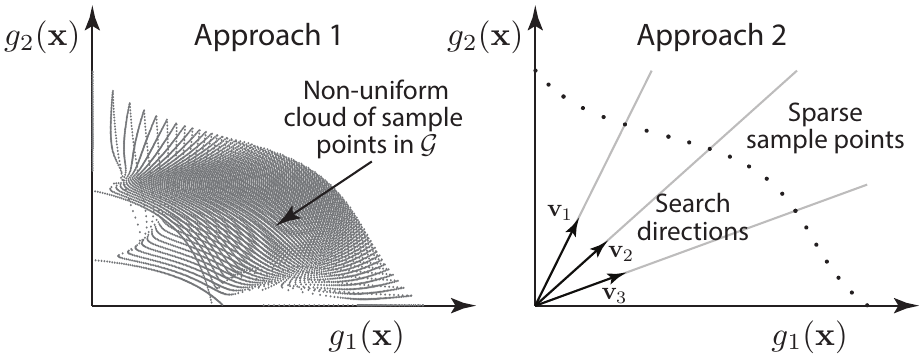} \vskip-4mm
	\caption{Illustration of the two approaches to visualize the objective set $\mathcal{G}$ by computing sample points.}
	\label{fig:visualization-example}
	\end{center} \vskip-8mm
\end{figure}

Fig.~\ref{fig:visualization-example} illustrates the two approaches. The first approach gives a cloud of sample points that provides a sense of the shape of $\mathcal{G}$: is it convex, what are the numerical ranges, and are the objectives strongly/weakly conflicting? The density of points is non-uniform and it is not guaranteed that any sample point is exactly on the Pareto boundary. In contrast, the second approach gives a sparse set of sample points that are exactly on the Pareto boundary. Each point is found by searching in a certain direction (e.g., $\vect{v}_1$, $\vect{v}_2$, $\vect{v}_3$) from the origin.

By looking at visualizations of the Pareto boundary, such as the ones in Fig.~\ref{fig:visualization-example}, the network designer can understand the fundamental properties and tradeoffs between conflicting objectives. Visualization is a powerful tool that supports the network designer in making an informed decision. This is the essence of the \emph{a posteriori} method. Since it is difficult to visualize more than three dimensions at a time, one needs to limit the granularity to a few objectives at a time. This issue can be treated in an iterative fashion where the network designer makes preliminary decisions (e.g., regarding the preferred minimal level for different objectives) which replaces the current resource bundle $\mathcal{X}$ with a smaller set $\tilde{\mathcal{X}} \subset \mathcal{X}$. This interactive process continues until the network designer is satisfied---a type of psychological convergence \cite{Branke2008a}.

\vspace{1mm}

\begin{mdframed}[linewidth=2pt]

\vspace{-4mm}

\section*{\small FINDING THE PARETO BOUNDARY BY BISECTION}

\vspace{-2mm}

The single-objective optimization problem in \eqref{eq:line-search-problem} finds the weak Pareto boundary in the direction $\vect{v}$ from the origin. This problem can be solved by checking if a series of points, each denoted  $\boldsymbol{\mu} = [\mu_1 \, \ldots \, \mu_M ]^T \in \mathbb{R}^M$,  belong to the attainable objective set $\mathcal{G}$ or not. This is determined by the \emph{membership test} \vspace{-2mm}
\begin{align} \label{eq:membership-test}
\mathrm{find} & \quad \vect{x} \in \mathcal{X} \\ \notag
\mathrm{subject} \, \mathrm{to} & \quad g_m(\vect{x}^*) =  \mu_m.
\end{align}

\vspace{-2mm}

\noindent The complexity of this feasibility problem is a baseline for other optimization problems that involve the same resource bundle and objective functions---if the membership test is computationally intractable, there is little chance that any meaningful problem formulation is practically solvable. Fortunately, there are many cases when the membership test is efficiently solvable; for example, it is a convex problem in many beamforming design problems for cellular networks \cite{Bjornson2013d}.

Equipped with a tractable membership test, we can solve \eqref{eq:line-search-problem} by first defining a range $[\lambda_{\min},\lambda_{\max}]$ of values for $\lambda$, such that $\lambda_{\min} \vect{v} \in \mathcal{G}$ and $\lambda_{\max} \vect{v} \not \in \mathcal{G}$. The lower limit can be $\lambda_{\min} = 0$, since the origin is always attainable. The upper limit is selected for the MOOP at hand, for example, by exploiting the utopia point (if it is known) or by relaxing the problem to find other unattainable points. The following algorithm solves \eqref{eq:line-search-problem}:

\begin{algorithmic}
\State \textbf{Input:} Range $[\lambda_{\min},\lambda_{\max}]$ and accuracy $\epsilon>0$
\While{$\lambda_{\max}-\lambda_{\min} > \epsilon$}
    \State Make membership test \eqref{eq:membership-test} for $\boldsymbol{\mu} = \frac{\lambda_{\max}+\lambda_{\min}}{2} \vect{v}$
    \If{ $\boldsymbol{\mu} \in \mathcal{G}$}
        \State $\lambda_{\min} \gets \frac{\lambda_{\max}+\lambda_{\min}}{2}$
    \Else
        \State  $\lambda_{\max} \gets \frac{\lambda_{\max}+\lambda_{\min}}{2}$
    \EndIf
\EndWhile
\State \textbf{Output:} Attainable point $\vect{a} = \lambda_{\min} \vect{v}$
\end{algorithmic}

 This is a classical bisection algorithm that cuts the range $[\lambda_{\min},\lambda_{\max}]$ in half in each iteration \cite{Boyd2004a}. Bisection has fast convergence and the distance between $\vect{a}$ and the Pareto boundary is below $\epsilon \|\vect{v}\|$ for given $\epsilon\!>\!0$.

\vspace{1mm}

\end{mdframed}

\vspace{-1mm}

\begin{figure}
	\begin{center}
		\includegraphics[width=\columnwidth]{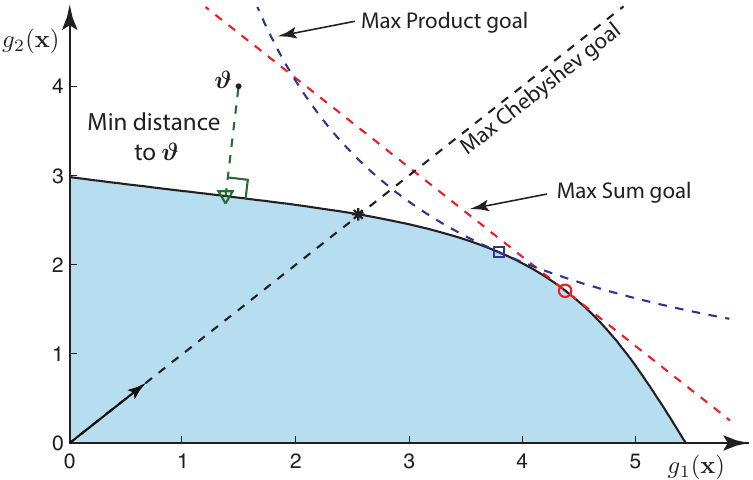} \vskip-3mm
	\caption{Illustration of the Pareto optimal operating points achieved by scalarization using common goal functions.}
	\label{fig:priori-example}
	\end{center} \vskip-7mm
\end{figure}

\subsection*{\small\em SOLVING A MOOP BY SCALARIZATION}

An alternative way to solve MOOPs in practice is the \emph{a priori method} where the network designer articulates preferences before any computations take place. The purpose is to find the operating point $\vect{g} \in \mathcal{G}$ that satisfies these preferences as well as possible. In particular, the designer can specify a \emph{goal function} $f: \mathbb{R}^{M} \rightarrow \mathbb{R}$ that for any conceivable operating point $\vect{g}$ (attainable or not) produces a scalar describing how preferable that point is (large value means high preference). The goal function describes a certain subjective tradeoff between the objectives and thus imposes an order on the vectors in the objective set $\mathcal{G}$. Consequently, the MOOP in \eqref{eq:optimization-problem} is converted into the single-objective optimization problem \vspace{-1mm}
\begin{align} \label{eq:scalarization}
\maximize{\vect{x}} & \quad f \big( g_1(\vect{x}), g_2(\vect{x}), \ldots, g_M(\vect{x}) \big) \\
\mathrm{subject} \, \mathrm{to} & \quad \vect{x} \in \mathcal{X}. \notag
\end{align}

 \vspace{-2mm}

\noindent This conversion is called \emph{scalarization} and the solution is a weak, and usually also strong, Pareto boundary point.  In contrast to the conventional approach of having a sole performance objective and expressing other potential objectives as constraints, \eqref{eq:scalarization} combines the $M$ objectives into a scalar goal function and has no additional constraints. It is indeed possible to impose constraints on the acceptable values for certain objectives also in the scalarization case, but it is not required.

The goal function can take many forms and a variety of classes of functions can be found in the literature; see \cite{Marler2004a,Branke2008a,Bot2009a,Bjornson2013d}.
We describe four important goal function classes. The most common goal function might be the weighted sum \vspace{-1mm}
\begin{equation} \label{eq:arthimetic}
f_{\mathrm{sum}}(\cdot) = \sum_{m=1}^{M} w_m g_m(\vect{x})
\end{equation}

 \vspace{-1mm}

\noindent where $w_1,\ldots,w_M$ are positive weights that specify the priority of each objective; the priority of the $m$th objective grows by increasing the corresponding weight $w_m$. One should be careful when interpreting the relative priorities, because the objectives can have different scales, units, and couplings.

Similarly, one can consider the weighted product \vskip-3mm
\begin{equation} \label{eq:geometric}
f_{\mathrm{product}}(\cdot) = \prod_{m=1}^{M} \big( g_m(\vect{x}) \big)^{w_m}
\end{equation}
where the weights are defined as before but act differently. Note that \eqref{eq:geometric} is the (weighted) geometric mean, while \eqref{eq:arthimetic} is the (weighted) arithmetic mean. Generally speaking, the geometric mean is better at comparing objectives with different numerical ranges, because the relative scaling has no impact.

The weighted Chebyshev formulation, also known as the weighted max-min formulation, played a key role when we computed sample points on the Pareto boundary in the \emph{a posteriori} method. The weighted Chebyshev goal function is
\begin{equation} \label{eq:chebyshev}
f_{\mathrm{chebyshev}}(\cdot) = \min_{1\leq m \leq M} \frac{ g_m(\vect{x})}{w_m}.
\end{equation}
This scalarization is equivalent to \eqref{eq:line-search-problem} if we write it on epigraph form \cite{Boyd2004a} and select the weights $w_1,\ldots,w_M$ as $w_m = v_m$ for all $m$. Hence, this scalarization searches for the Pareto boundary in the direction $[w_1 \, \ldots \, w_M]^T$ from the origin.

Alternatively, the network designer can specify a preferable operating point $\boldsymbol{\vartheta} \in \mathbb{R}^{M}$ (e.g., the utopia point $\boldsymbol{\vartheta} = \vect{u}_{\mathrm{utopia}}$). The \emph{distance goal function} is defined as \vspace{-2mm}
\begin{equation} \label{eq:goaldistance}
f_{\mathrm{distance}}(\cdot) = - \| \boldsymbol{\vartheta} - \vect{g}(\vect{x}) \|
\end{equation}
and measures the distance from the preferable point in some appropriately selected norm $\|\cdot\|$. The norm $\| \boldsymbol{\vartheta} - \vect{g}(\vect{x}) \|$ should be small (preferably zero), thus the negative sign is used to achieve a goal function that is to be maximized.

The final operating point is determined by the choice of goal function. Interestingly, the computational complexity also varies with the goal function; the scalarized problem in \eqref{eq:scalarization} may be convex (i.e., solvable in polynomial time) for some classes of functions, while other classes give non-convex problems with exponential complexity---or even worse. For example, \cite{Liu2011a} proved that transmit beamforming optimization in cellular networks is (quasi-)convex for the weighted Chebyshev goal function and strongly NP-hard for most other goal functions. This result has general implications.

\begin{property} \label{property:pragamatic-approach}
The weighted Chebyshev goal function is the safest choice in terms of computational complexity; if there exists a tractable membership test, it can be solved efficiently as described in ``Finding the Pareto Boundary by Bisection''.
\end{property}

Since goal functions are inherently subjective, no choice is better than the others in terms of optimality. Property \ref{property:pragamatic-approach} inspired \cite{Bjornson2013d} to propose what is known as \emph{the pragmatic approach to resource allocation}: select the weighted Chebyshev goal function (due to its tractable complexity) and exploit the weights to adapt to the needs of the network designer.

The operating points attained by different scalarizations are illustrated in Fig.~\ref{fig:priori-example}, for a scenario where the attainable ranges are different for the two objectives. The goal functions in \eqref{eq:arthimetic}--\eqref{eq:goaldistance} are considered for $w_1=w_2=1$. Let $f^*$ denotes the optimal function value in \eqref{eq:scalarization}, which of course is different for each goal function. The optimal operating point with the sum goal function lies on the level curve $f_{\mathrm{sum}}(\cdot) = f^*$, which is the red line in Fig.~\ref{fig:priori-example}. Similarly, $f_{\mathrm{product}}(\cdot) = f^*$ gives the blue parabolic level curve of the product goal function. These level curves touch the objective set $\mathcal{G}$ in unique Pareto boundary points, which are the optimal operating points for the respective scalarized problems. As described earlier, the Chebyshev goal function searches on a line from the origin. For $w_1=w_2=1$ this is the line where the two objectives have equal values. If there is a preferable operating point $\boldsymbol{\vartheta} \not \in \mathcal{G}$ as in  Fig.~\ref{fig:priori-example}, \eqref{eq:scalarization} provides the operating point that minimizes the distance to $\mathcal{G}$ (the Euclidean distance is used in Fig.~\ref{fig:priori-example}).

The function classes in \eqref{eq:arthimetic}--\eqref{eq:chebyshev} are parameterized by the weights $\vect{w}= [w_1,\ldots,w_M]^T$. Different weight selections give different Pareto optimal points when solving \eqref{eq:scalarization}. By varying $\vect{w}$ over the set $\mathcal{W} = \{ \vect{w} : w_m \geq 0 \,\, \forall m, \,\, \sum_m w_m = 1 \} \in \mathbb{R}^{M}$ of positive weights that sum up to one, we can attain the whole Pareto boundary or a subset thereof, depending on the function class \cite{Branke2008a}. Since each scalarization in \eqref{eq:scalarization} is a single-objective optimization problem, it is equipped with conventional Karush-Kuhn-Tucker (KKT) optimality conditions \cite{Boyd2004a}. By considering all $\vect{w} \in \mathcal{W}$, these can be extended to a joint set of optimality conditions for all points achieved by the function class \cite{Bot2009a}. These optimality conditions describe the structure of the resource utilizations that achieve the Pareto boundary; for example, it was utilized in \cite[Section 3.2]{Bjornson2013d} to parameterize any efficient transmit beamforming.

Finally, we note that game theory provides an alternative way to select operating points from the Pareto boundary, by specifying the rules of a game instead of a goal function \cite{Jorswieck2013a}. These techniques are mainly for systems with separate objectives that compete for shared resources, while single-operator networks typically have dedicated resources.

\vspace{-3mm}

\section*{\small CASE STUDY: DESIGNING MASSIVE MIMO SYSTEMS}

\vspace{-2mm}

We exemplify the usefulness of MOO by a case study. The goal is to visualize tradeoffs between conflicting 5G objectives and describe how the framework can be used to acquire new insights and prove old heuristic observations.  In recent years, coordinated multipoint (CoMP) techniques have shown the potential to greatly improve the area rates in cellular networks. This is achieved by deploying antenna arrays at base stations (BSs) and apply a coordinated space division multiple access (SDMA) scheme across the network \cite{EP0616742,Gesbert2007a,Gesbert2010a,Bjornson2013d}. Unfortunately, CoMP is difficult to implement since the coordination signaling is limited \cite{Marsch2008a}, the signal processing complexity increases drastically \cite{Liu2011a}, and the performance gains are not robust to the inter-user interference caused by having imperfect channel state information (CSI) \cite{Gesbert2007a}.

The concept of massive MIMO has gained traction since it might eliminate the CoMP issues listed above  \cite{Marzetta2010a,Rusek2013a,Hoydis2013c,Bjornson2014a}. Massive MIMO is based on the idea of deploying large arrays with unconventionally many active antennas at the BSs and serve a much smaller number of users; for example, hundreds of antennas that serve several tens of users. One would imagine that adding more antennas and users into a system would make CoMP even more difficult to implement, but the beauty of massive MIMO is that this is not the case \cite{Marzetta2010a}. The excessive number of antennas brings robustness to imperfect CSI, makes low-complexity signal processing close to optimal \cite{Rusek2013a}, and allows for simple implicit intercell coordination \cite{Hoydis2013c}. Massive MIMO systems are even robust to the distortions caused by hardware imperfections \cite{Bjornson2014a}.

In this case study, we strive to optimize the downlink transmission of a massive MIMO system to balance $M=3$ conflicting objectives: high average user rates, high average area rates, and high energy efficiency. The cellular network that we consider has 16 cells, each consisting of a BS with $N$ antennas and $K$ single-antenna users. The bandwidth is $B = 10 \, \mathrm{MHz}$, the emitted power per BS is denoted $P$ Watt ($\mathrm{W}$), and $\sigma^2 = 10^{-13} \, \mathrm{W}$ is the average noise power.

\begin{figure}
	\begin{center}
		\includegraphics[width=\columnwidth]{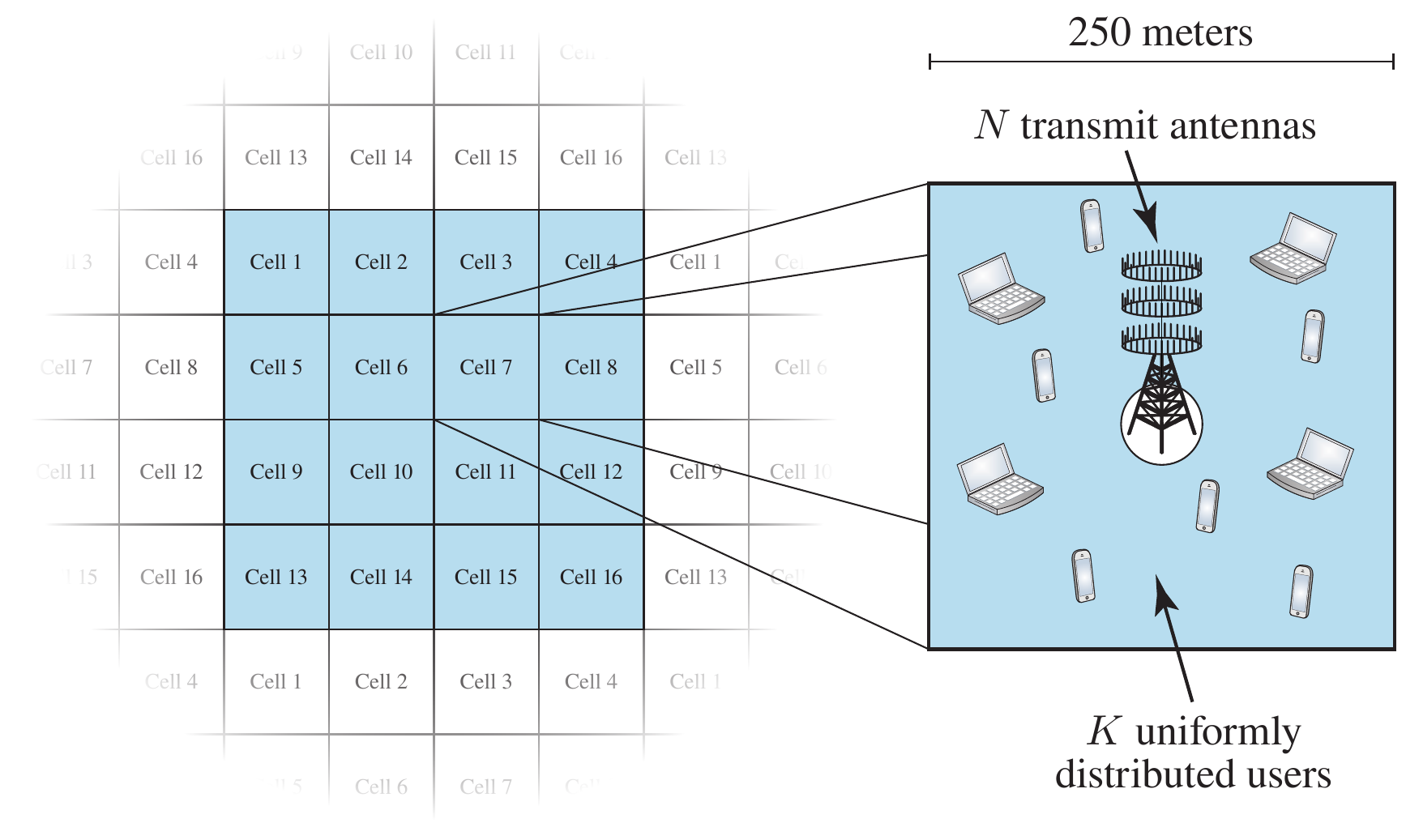} \vskip-4mm
	\caption{Illustration of the scenario in the case study: A cellular network with $N$ antennas per BS and $K$ users per cell.}
	\label{figure:simulationscenario}
	\end{center} \vskip-5mm
\end{figure}

Each cell is a square of size $250 \times  250$ meters (i.e., the area is $A = 0.25^2 \, \mathrm{km}^2$) and we apply classic wrap-around to avoid edge effects; the scenario is shown in Fig.~\ref{figure:simulationscenario}. The $K$ users are uniformly distributed in the cell, with a minimum distance of $35$ meters. For a randomly picked user, let $\lambda_{\textrm{serving}}$ be the channel variance from the serving BS and $P \lambda_{\textrm{intercell}}$ be the average intercell interference power. We are concerned with average behaviors and define the expectations $\Lambda_1 = \mathbb{E}\{ \frac{1}{\lambda_{\textrm{serving}}} \}$ and $\Lambda_2 = \mathbb{E}\{ \frac{\lambda_{\textrm{intercell}}}{\lambda_{\textrm{serving}}} \}$ for later use.
Using the same 3GPP pathloss model as in \cite{Bjornson2015a}, we get $\Lambda_1 = 1.72 \cdot 10^9$ and $\Lambda_2 = 0.54$.

The optimization/resource variables in this case study are the number of BS antennas $N$, the number of users $K$, and the transmit power $P$ per cell. The resource bundle is
\begin{equation} \label{eq:casestudy:resourceset}
\mathcal{X} = \left\{
[K \,\, N \,\, P]^T  :
\begin{array}{l}
1 \leq K \leq \frac{N}{2}, \\
2 \leq N \leq N_{\max}, \\
0 \leq P \leq N P_{\max}
\end{array} \!\!
\right\}
\end{equation}
where $N_{\max}=500$ is the maximal number of antennas that can fit at each BS, $P_{\max} = 20  \, \mathrm{W}$ is the maximal emitted power per BS antenna, and the constraint $K \leq \frac{N}{2}$ makes sure that we have many more BS antennas than active users.

Next, we define the average user rate and the total power consumption per cell. For simplicity, we assume that each BS has obtained perfect CSI for its users and applies zero-forcing precoding, which is a signal processing technique that cancels out intracell interference by beamforming and adapts the power allocation to guarantee the same rate to each user. Similar to \cite{Bjornson2015a}, the average user rate can be shown to be
\begin{equation} \label{eq:rate-expression}
R_{\textrm{average}} = B \left( 1 - \frac{K}{\Upsilon} \right) \log_2 \left( 1 + \frac{\frac{P}{K}(N-K)}{\sigma^2 \Lambda_1 + P \Lambda_2} \right),
\end{equation}
under the assumption that each user knows its useful channel and treats intercell interference as noise. The prelog-factor
$( 1 - \frac{K}{\Upsilon} )$ accounts for the necessary overhead for channel acquisition and $\Upsilon = 1000$ is the number of channel uses that the channel stays fixed. It is selected as $\Upsilon = B_{\textrm{coherence}} \tau_{\textrm{coherence}}$, where $B_{\textrm{coherence}} = 200 \, \mathrm{kHz}$ is the coherence bandwidth and $\tau_{\textrm{coherence}} = 5 \, \mathrm{ms}$ is the coherence time. Looking inside the logarithm of \eqref{eq:rate-expression}, $\frac{P}{K}$ is the average transmit power per user, $N-K$ is the effective array gain, and $\sigma^2 \Lambda_1 + P \Lambda_2$ is the average degradation from noise and intercell interference.

Based on the models and the practical numbers in \cite{Bjornson2015a,Yang2013a,Auer2011a}, the total power consumption per cell is given by
\begin{equation}
P_{\textrm{total}} = \frac{P}{\eta} + N C_N + K C_K + \frac{C_{\textrm{precoding}}}{L} + C_0
\end{equation}
where $\eta = 0.31$ is the efficiency of the power amplifiers at the BS, $C_N = 0.5 \, \mathrm{W}$ is the hardware power consumed per transmit antenna, $C_K = 0.3 \, \mathrm{W}$ is the hardware power per user, and $C_0 = 10 \, \mathrm{W}$ is the static hardware power. In addition, $C_{\textrm{precoding}} = 3 K^2 N \frac{B}{\Upsilon}$ is the floating-point operations per second (flops) required to compute zero-forcing precoding, while $L = 12.8 \, \mathrm{Gflops/W}$ is a typical computational efficiency.

We are now ready to define our three objective functions: \vspace{-2mm}
\begin{align} \label{eq:g1}
g_1(\vect{x}) &= R_{\textrm{average}} & [\mathrm{bit}/\mathrm{s}/\mathrm{user}]\\ \label{eq:g2}
g_2(\vect{x}) &= \frac{K}{A} R_{\textrm{average}} &  [\mathrm{bit}/\mathrm{s}/\mathrm{km}^2]\\ \label{eq:g3}
g_3(\vect{x}) &= \frac{K R_{\textrm{average}}}{P_{\textrm{total}}} & [\mathrm{bit}/\mathrm{J}]
\end{align}
where $\vect{x}=[K \, N \, P]^T$ are the optimization/resource variables. The objective $g_1(\vect{x})$ is the average user rate, $g_2(\vect{x})$ is the average area rate, and $g_3(\vect{x})$ is the energy efficiency.

\vspace{-3mm}

\subsection*{\small\em DESIGNING MASSIVE MIMO BY MOO FRAMEWORK}

\vspace{-2mm}

We have now defined a MOOP of the type in \eqref{eq:optimization-problem}. The resource bundle is given by \eqref{eq:casestudy:resourceset} and the three objectives are defined in \eqref{eq:g1}--\eqref{eq:g3}. We now describe how the MOO framework can be used to study tradeoffs between these objectives,  with the purpose of deriving new insights and confirming old beliefs.

\begin{figure}[ht!]
	\begin{center}
		\includegraphics[width=\columnwidth]{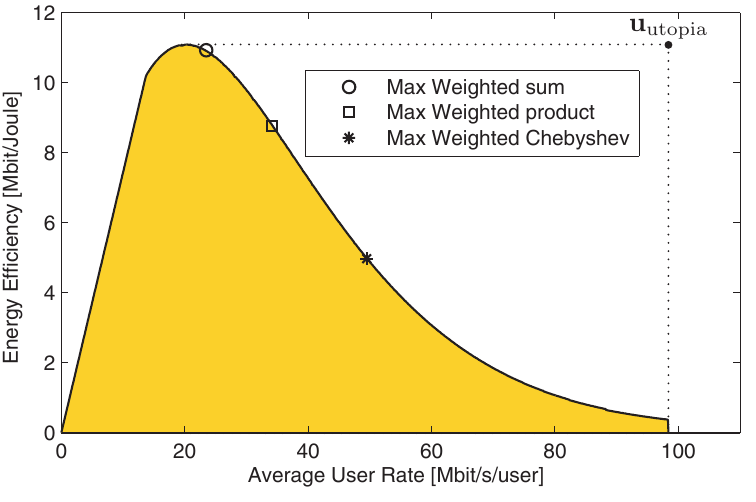} \vskip-4mm
	\caption{Visualization of the tradeoff between two objectives in the case study: average user rate and energy efficiency.}
	\label{figure:casestudy2dSE}
	\end{center} \vskip-6mm
\end{figure}

\begin{figure}[ht!]
	\begin{center}
		\includegraphics[width=\columnwidth]{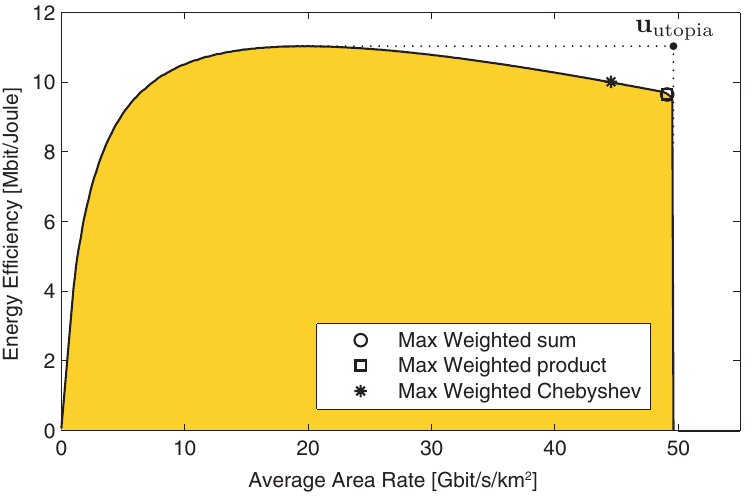} \vskip-4mm
	\caption{Visualization of the tradeoff between two objectives in the case study: average area rate and energy efficiency.}
	\label{figure:casestudy2dASE}
	\end{center} \vskip-6mm
\end{figure}

The tradeoff between the average user rate and the EE is shown in Fig.~\ref{figure:casestudy2dSE}. The objective set with respect to these two objectives was generated by the second approach described earlier (i.e., searching for the Pareto boundary in different directions). Fig.~\ref{figure:casestudy2dSE} shows that these two objectives are aligned up to the point $g_1 = 20.4 \, \mathrm{Mbit}/\mathrm{s}/\mathrm{user}$ and $g_3 = 11.1 \, \mathrm{Mbit}/\mathrm{J}$, where the maximal EE is achieved. The objectives are then conflicting, because the user rates can only be further increased by making drastic sacrifices in the EE.

\begin{figure}[t!]
	\begin{center}
		\includegraphics[width=\columnwidth]{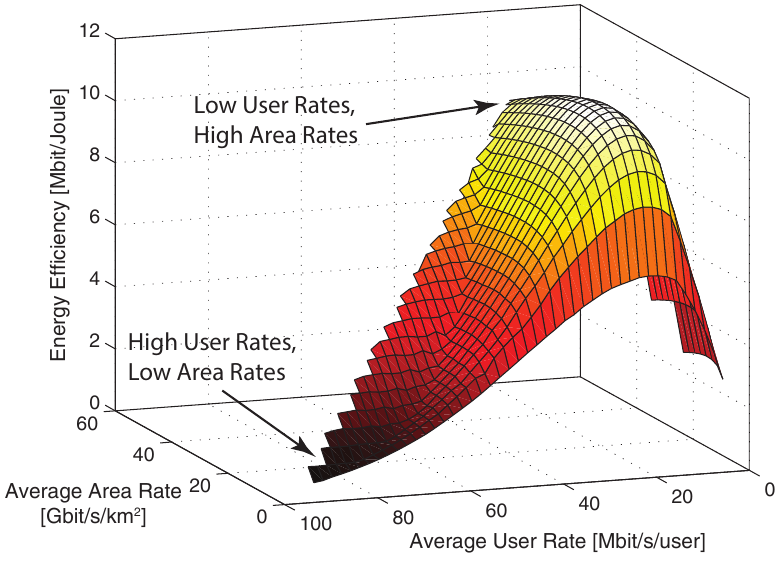} \vskip-4mm
	\caption{Visualization of the tradeoff between all three objectives in the case study: average user rate, average area rate, and energy efficiency.}
	\label{figure:casestudy3d}
	\end{center} \vskip-8mm
\end{figure}

Another tradeoff is illustrated in Fig.~\ref{figure:casestudy2dASE}, where the average area rate and the EE are compared. These objectives are also aligned until the EE reaches its maximum value. However, one can increase the area rate beyond this point with only minor losses in EE. By noting that $ g_2(\vect{x}) = \frac{K}{A} g_1(\vect{x})$ and comparing with the previous figure, this obviously means that the area rate is improved by transmitting to more users (i.e., having a larger $K$) and not by increasing the rate per user. This conclusion is supported by Fig.~\ref{figure:casestudy3d}, which shows the three-dimensional objective set with respect to all objectives.

Fig.~\ref{figure:casestudy3d} reveals that high area rates are only achievable when the rate per active user is low, which means that we serve many user devices in parallel. In contrast, high rates per user is only achievable by having fewer active users. High energy efficiency is possible when the rate per user is small. These different operating points are achieved by different resource utilizations $\vect{x} \in \mathcal{X}$; thus, the number of antennas/users are different and the signal processing related to precoding changes. This proves the otherwise heuristic belief that the network architecture must be flexible (e.g., in terms of switching off antennas and precoding adaptation) if different operating points should be attainable in different traffic cases.

The discussion above is typical for the \emph{a posteriori} method; we analyzed the shape of the Pareto boundary and drew conclusions on which operating points that are preferable to us. If we would instead utilize the \emph{a priori} method, then we need to specify a goal function. This can be done by picking any of the function classes described in \eqref{eq:arthimetic}--\eqref{eq:goaldistance} and selecting the corresponding parameters (e.g., weights) to describe our subjective goals. To aid us in this process, suppose we know the utopia point $\vect{u}_{\mathrm{utopia}}$ (defined in \eqref{eq:utopiapoint}) in advance. This point contains the maximal value for each objective, if we would focus completely on it. If the three objectives are equally important to us, it makes sense to normalize their numerical ranges. This is achieved by setting $\vect{w} = [\frac{1}{u_1} \,\, \frac{1}{u_2} \,\, \frac{1}{u_3}]^T$ in the weighted sum goal, $\vect{w} = [1 \,\, 1 \,\, 1]^T$ in the weight product goal, and $\vect{w} = \vect{u}_{\mathrm{utopia}}$ in the weighted Chebyshev goal formulation. The corresponding operating points when solving the scalarized problem in \eqref{eq:scalarization} are shown in Fig.~\ref{figure:casestudy2dSE} and Fig.~\ref{figure:casestudy2dASE}. The shape of the region has a great impact on the spread of the operating points, but different weights still give different operating points (as discussed earlier). The utopia point $\vect{u}_{\mathrm{utopia}} = [u_1 \, u_2 \, u_3]^T$ is also shown in these figures. We observe that it is far outside the attainable objective set in Fig.~\ref{figure:casestudy2dSE}, since the two objectives are strongly conflicting. On the contrary, the utopia point is quite close to the objective set in Fig.~\ref{figure:casestudy2dASE}, where the conflict is rather mild.

Finally, we remark that the \emph{a posteriori} and \emph{a priori} methods can be combined. The network architecture can, for example, be designed by studying the shape of the attainable objective set and making sure that the network can adapt and achieve different operating points at the Pareto boundary at different times. The system designer can then formulate multiple goal functions that are exploited for efficient real-time network adaptation, based on current traffic load, service requirements, and capability of the user devices.

\section*{\small CONCLUSIONS AND FUTURE DIRECTIONS}

\vspace{-2mm}

The design expectations on 5G wireless networks cannot be properly articulated by a single performance objective. There are many conflicting objectives, such as improving the peak user rates, average area rates, and energy efficiency. The network design thus calls for multi-objective optimization, which is rigorous framework for studying and solving design problems with multiple objectives. This article provided a survey on this topic. There is no objectively optimal solution to this type of problems, but there are two main methods to find subjectively optimal solutions that fit the needs of the network designer. The \emph{a posterior method} computes sample points on the Pareto boundary---the set of tentative operating points where no objective can be improved without degrading another objective. The sample points are used to visualize the Pareto boundary for the network designer, who can then make well-informed design decisions. Alternatively, the network designer can specify a goal function that describes the acceptable tradeoffs between objectives and infers an order on the attainable operating points. One can then maximize this tradeoff by solving a conventional optimization problem and thereby obtain the most suitable Pareto boundary point.

We also provided a case study on network dimensioning of cellular networks that allows for massive MIMO deployment. This example illustrates our vision of how the MOO framework can be utilized to balance conflicting performance objectives when designing future wireless communication networks. While the analytic tools provided by MOO are well-established, the applications to communication networks are greatly unexplored.  A particular research challenge is to formulate MOOPs with a modeling granularity that allows us to answer fundamental design questions related to how the system can efficiently manage the heterogeneous 5G characteristics described in the introduction. To this end, the models must capture the main practical propagation characteristics, be robust to hardware imperfections and uncertain model parameters, and allow for optimization of the signal processing techniques. All of this is to be done while making the basic optimization operations (e.g., the membership test described above) computationally tractable.

\vspace{-3mm}

\section*{\small ACKNOWLEDGMENTS}

\vspace{-2mm}

This article has been supported by the International Postdoc Grant 2012-228 from the Swedish Research Council and the ERC Starting Grant 305123 MORE (Advanced Mathematical Tools for Complex Network Engineering).

\vspace{-3mm}

\section*{\small AUTHORS}

\vspace{-2mm}

{\bf\em Emil Bj\"ornson} (emil.bjornson@liu.se) received the M.S. degree from Lund University, Sweden, in 2007 and the Ph.D. degree in 2011 from KTH Royal Institute of Technology, Sweden. From 2012 to 2014, he was a joint postdoctoral researcher at Sup\'{e}lec, France, and KTH Royal Institute of Technology, Sweden, sponsored by a personal International Postdoc Grant from the Swedish Research Council. He is the first author of the book ``Optimal Resource Allocation in Coordinated Multi-Cell Systems'' and received best conference paper awards in 2009, 2011, and 2014. He is now a Research Fellow in the tenure-track at Link\"{o}ping University, Sweden.

{\bf\em Eduard Jorswieck} (eduard.jorswieck@tu-dresden.de) received the M.S. and Ph.D. degrees from the Technische Universit\"{a}t Berlin, Germany, in 2000 and 2004, respectively.
He was with the Fraunhofer Institute for Telecommunications, Heinrich-Hertz-Institut (HHI) Berlin, from 2000 to 2008. From 2005 to 2008, he was a lecturer at the Technische Universit\"{a}t Berlin. From 2006 to 2008 he was a post-doc and assistant professor at the KTH Royal Institute of Technology, Sweden. Since 2008, he has been the head of the Chair of Communications Theory and Full Professor at Dresden University of Technology (TUD), Germany. In 2006, he received the IEEE Signal Processing Society Best Paper Award.

{\bf\em M\'erouane Debbah} (merouane.debbah@supelec.fr) received the M.S. and Ph.D. degrees from \'{E}cole Normale
Sup\'{e}rieure de Cachan, France, in 1999 and 2002, respectively. From 1999 to 2002, he worked for Motorola Labs. From 2002 to 2003, he was appointed senior researcher at the Vienna Research Center for Telecommunications (FTW) (Austria). From 2003 to 2007, he joined the mobile communications department of the Institut Eurecom, France, as an assistant professor. He is currently a professor at Sup{\'e}lec, France, and holds the Alcatel-Lucent Chair on Flexible Radio.  He received the 2005 Mario Boella Prize Award, the 2007 GLOBECOM Best Paper Award, the 2009 Wi-Opt Best Paper Award, the 2010 Newcom++ Best Paper Award, as well as the 2007 Valuetools, 2008 Valuetools, and 2009 CrownCom Best Student Paper Awards. He is a WWRF fellow.

{\bf\em Bj\"orn Ottersten} (bjorn.ottersten@uni.lu) received the M.S. degree from Link\"{o}ping University, Sweden, in 1986 and the Ph.D. degree in 1989 from Stanford University, California. In 1991 he was appointed Professor of Signal Processing at KTH Royal Institute of Technology, Sweden.  During 96/97 Dr. Ottersten was Director of Research at ArrayComm Inc, a start-up based on Ottersten's patented technology. Currently, Dr.~Ottersten is Director for the Interdisciplinary Centre for Security, Reliability and Trust at the University of Luxembourg. He coauthored articles that received the IEEE Signal Processing Society Best Paper Award in 1993, 2001, 2006, and 2013, and several IEEE conference papers receiving Best Paper Awards. In 2011 he received the IEEE Signal Processing Society Technical Achievement Award. He is editor-in-chief of EURASIP Signal Processing Journal. He is a Fellow of the IEEE and EURASIP.

\vspace{-3mm}

\section*{\small REFERENCES}

\bibliographystyle{IEEEbib}
\bibliography{IEEEabrv,refs}

\enlargethispage{5mm}

\end{document}